%% file: main.tex
\documentclass{article}

\usepackage{arxiv}

\usepackage[utf8]{inputenc} 
\usepackage[T1]{fontenc}    
\usepackage{hyperref}       
\usepackage{url}            
\usepackage{booktabs}       
\usepackage{amsfonts}       
\usepackage{nicefrac}       
\usepackage{microtype}      
\usepackage{lipsum}
\usepackage{graphicx}
\usepackage{amsmath}
\usepackage{comment}
\usepackage{subfiles} 
\graphicspath{ {figs/} }

\title{Face-to-Music Translation Using a Distance-Preserving Generative Adversarial Network with an Auxiliary Discriminator}

\author{
 Chelhwon Kim \\
  FX Palo Alto Laboratory\\
  Palo Alto, CA 94304 \\
  \texttt{chkim@soe.ucsc.edu} \\
   \And 
 Andrew Port \\
  Department of Computer Engineering\\
  University of California, Santa Cruz\\
  Santa Cruz, CA 95064 \\
  \texttt{aaport@ucsc.edu} \\
  \And
 Mitesh Patel \\
  FX Palo Alto Laboratory\\
  Palo Alto, CA 94304 \\
  \texttt{mitesh@fxpal.com} \\
}

\DeclareUnicodeCharacter{2212}{-}

\begin{document}
\maketitle

\newcommand\videodemoaddress{\url{https://www.dropbox.com/s/the176w9obq8465/face_to_musical_note.mov?dl=0}}

\newcommand\appendixsection{the appendix section}

\newcommand\x{3.0 in}

\newcommand\y{5in}

\begin{abstract}
Learning a mapping between two unrelated domains-such as image and audio, without any supervision is a challenging task. In this work, we propose a distance-preserving generative adversarial model to translate images of human faces into an audio domain. The audio domain is defined by a collection of musical note sounds recorded by 10 different instrument families (NSynth~\cite{nsynth2017}) and a distance metric where the instrument family class information is incorporated together with a mel-frequency cepstral coefficients (MFCCs) feature. To enforce distance-preservation, a loss term that penalizes difference between pairwise distances of the faces and the translated audio samples is used. Further, we discover that the distance preservation constraint in the generative adversarial model leads to reduced diversity in the translated audio samples, and propose the use of an auxiliary discriminator to enhance the diversity of the translations while using the distance preservation constraint. We also provide a visual demonstration of the results and numerical analysis of the fidelity of the translations. A video demo of our proposed model's learned translation is available in \videodemoaddress.
\end{abstract}



\subfile{body}

\section{Appendix}
\subfile{appendix}

\bibliographystyle{unsrt}
\bibliography{main}

\end{document}

%% file: body.tex
\section{Introduction}
There has been a lot of work in attempting to find a meaningful mapping between two different domains without any supervision - i.e. only unordered and unpaired samples from the two domains are given. However, most of them are often limited to a task of mapping between two visual domains such as image-to-image translation~\cite{gatys2016image, johnson2016perceptual, isola2017image, CycleGAN2017}, where the learner takes an image in one domain and maps it into another pixel domain with its content or style changed to be similar to that of the target samples. Another limitation is that it is often difficult to find a good use case for the image-to-image translation in real-world applications. In this work, we focus on finding a mapping between two highly unrelated domains - image and audio. More precisely, we want to find a meaningful mapping that links images of faces into musical sounds, while the measure of similarity between face images is consistent with the one between translated audio samples so that visually dis/similar face images can be mapped into audio samples that sound perceptually dis/similar to each other. This could be used to aid those with a visual impairment by encoding the visual information (e.g. facial appearance) into the audio domain, which allows them to perceive this information using their ears. In a recent work by~\cite{port2020earballs}, a neural transmodal translation system that translates images of faces into human speech-like audio clips was proposed, and was proven to be useful in a human subject test, where participants were able to successfully classify images of faces while only listening to the translated face audio samples. 

Finding a meaningful mapping between such unrelated domains is a challenging task. Recent deep learning-based approaches~\cite{CycleGAN2017} tackle this problem by learning a mapping simultaneously with its inverse mapping using a cycle-consistency loss and showed that this allows to obtain a convincing mapping between two visual domains such as images of horses to zebras, summer to winter, etc. The more recent approach of \cite{port2020earballs} has shown that a distance preserving constraint~\cite{onesided} in the generative adversarial learning framework can learn a meaningful map from faces images to human speech-like audio clips. This approach first leverages a pretrained image embedding network to map images into a compact subset of Euclidean space, where $L^2$ distance corresponds to a measure of similarity. Then, a generative adversarial network (GAN) is trained to find a distance preserving map from the image embedding space into a new audio metric space defined by a collections of target audio samples equipped with a mel-frequency cepstral coefficients (MFCCs)-based distance metric. To enforce the distance preservation, a loss term that penalizes differences between the pairwise distances of samples in each input batch and the pairwise distances of samples in corresponding batch of generated audio samples output.  

In this work, we adapt the distance-preserving technique to map images of faces into an audio domain defined by a collection of musical note sounds recorded by 10 different instrument families (NSynth~\cite{nsynth2017}), and we design a new similarity measure of those audio samples where the instrument family class information is incorporated together with the MFCCs, as a result, musical samples of the same instrument family with similar timbre have small distances and musical samples of different instruments with dissimilar timbre have large distances. With this, we believe that the system can produce sounds that are easier to discern for humans and whose difference are more memorable. 
Further, as we'll discuss in our experimental result section, 
our model also demonstrates that, when translating between such unrelated modalities, there is a trade-off between the variety of the translations (i.e. audio), and the preservation of geometric information. 
To address this problem, we propose using an auxiliary discriminator that can support the main discriminator in a way that it further enforces that the model outputs sounds which fit into the target audio dataset in the customized audio metric space. 

Contributions of our work can be summarized as follows.
\begin{itemize}
    \item We propose a distance preserved generative adversarial model 
    to learn a mapping from face images into a new audio domain that is defined by a collection of musical note sounds (NSynth dataset).
    \item We design a new audio metric customized for the NSynth dataset where their annotated instrument class labels are incorporated together with the MFCCs features. 
    \item We provide numerical analysis and further visual (T-SNE) results showing that facial information is being preserved and also that the produced sounds fit into the target dataset (according to FID).
    \item We discover that applying the distance preservation constraint to a generative model can lead to reduced diversity of the generated audio and demonstrate a method of using an additional discriminator to restore and enhance output diversity. 
\end{itemize}

\subsection{Related Work}
\subsubsection{GAN-based audio synthesis}
GANs have seen great success in generating images with high-fidelity~\cite{brock2018large, karras2017progressive, miyato2018spectral}, however, directly adapting the image generation to the audio domain fails to get a similar level of fidelity. Recent work~\cite{engel2019gansynth} discovered that generating log-magnitude spectrograms (i.e. time-frequency representation of audio) and phases directly with a progressively trained GAN~\cite{karras2017progressive} can produce more coherent waveforms. While operating GANs on the image like spectrograms for audio generation showed promising results, \cite{wavegan} investigated a different method that applies the GAN to raw-waveform audio generation. This method uses a one-dimensional deep convolutional GAN (DCGAN)~\cite{radford2015unsupervised} with longer, one-dimensional filters and a larger upsampling factor to capture structure across a long range of timescales, and showed that it can successfully synthesize audio from many different domains such as human speech, drums, bird vocalizations, and piano. Our model is built on top of this model to directly generate the raw-waveform audio.

There also have been approaches to audio synthesis using autoregressive generative models~\cite{oord2016wavenet, mehri2016samplernn} that predict each audio sample conditioned on the samples at the previous timestamps. 
However, these models suffer from their slow generation since the predictions are sequential-i.e. after each sample is predicted, it is fed back into the network to predict the next sample. GAN-based approaches have demonstrated advantages over these methods since the GAN can generate the entire audio sample with a single forward generative network. 

\subsubsection{Unsupervised mapping techniques}
There is significant research interest in unsupervised translation, especially with regard to image-to-image mapping.  CycleGAN \cite{CycleGAN2017} and DiscoGAN \cite{kim2017learning} are two examples of models which seek learn a map to convert images from a first dataset into images which appear to convincingly come from a second dataset.  Both these models simultaneously learn this map, $f$, and inverse map, $g$, such that $f(g(x)) \approx x$ and $g(f(y)) \approx y$ for images $x$ and $y$ from the first and second dataset respectively.  For example, this can learn a mapping from a dataset of photos taken in winter to a dataset of photos taken in summer.  A later work, \cite{onesided}, was able to produce similar results to \cite{CycleGAN2017} and \cite{kim2017learning} without learning the inverse map, $g$, by enforcing that $f$ was distance preserving with respect a pairwise distance between images in a batch.  Further works have also demonstrated that more specific information can be preserved in such a translation, e.g. \cite{kaur_eyegan_2020} used this method to augment a gaze-recognition dataset.  Their model enforced that $f$ commute with a pre-trained segmentation model, $s$, (i.e. that $f\circ s = s\circ f$).  In this way they were able to produce new images similar to those in their gaze-recognition dataset while preserving the accuracy of their ground truth gaze-direction labels.

\subsubsection{Image to audio mapping}
\cite{tian2019latent} proposed a method that finds a shared ``bridging" variational autoencoder (VAE)~\cite{kingma2013auto} network that maps pretrained latent representations of samples from each domain (i.e. image and audio) into a shared latent space while 
encouraging the distributions of the samples from the two domains are overlapped each other by minimizing their sliced-wasserstein distance (SWD)~\cite{bonneel2015sliced}. 
As the distance computation does not require the paired samples, this method finds the mapping in the unsupervised fashion, however, to further impose a semantic alignment between two domains, a linear classifier was used to separate the samples in the shared latent space by their annotated class-labels. More recent approach \cite{port2020earballs} and closest to our work, has shown that a distance preserving constraint~\cite{onesided} in the generative adversarial learning framework finds a meaningful map from faces images to human speech-like audio clips, and in human subject tests, users were able to accurately classify audio translations
of faces. Our method adapts the similar approach but focused on generating musical notes from faces 
and designed a new audio metric where both the discrete class labels and the MFCC-based distance were taken into account. 

\subsubsection{Variational autoencoder}
The variational autoencoder~\cite{kingma2013auto} has been shown that it learns a continuous and smooth latent space, where the samples generated from this space captures all the variations in the training samples and the variation in the latent space causes a smooth transformation between generated samples. Some recent work~\cite{engel2017latent, makhzani2015adversarial} has focused on imposing a specified prior distribution on the latent vectors and as a result, this learns a generative model-i.e. the encoder part of the autoencoder that maps imposed prior to given data distribution. As we will show in our experimental results, we tested an adversarial autoencoder by~\cite{makhzani2015adversarial} to impose a distribution of the face images on the latent space and used the decoder to generate the audio samples. This method employs the adversarial learning method to impose the prior distribution on the latent distribution by training a discriminative network that is attached on top of the latent vector to distinguish samples from the prior distribution from samples generated by the encoder network while the encoder is trained to fool the discriminator with its generated samples.

\section{Method}
\begin{figure}
	\centering
	\includegraphics[width=4in]{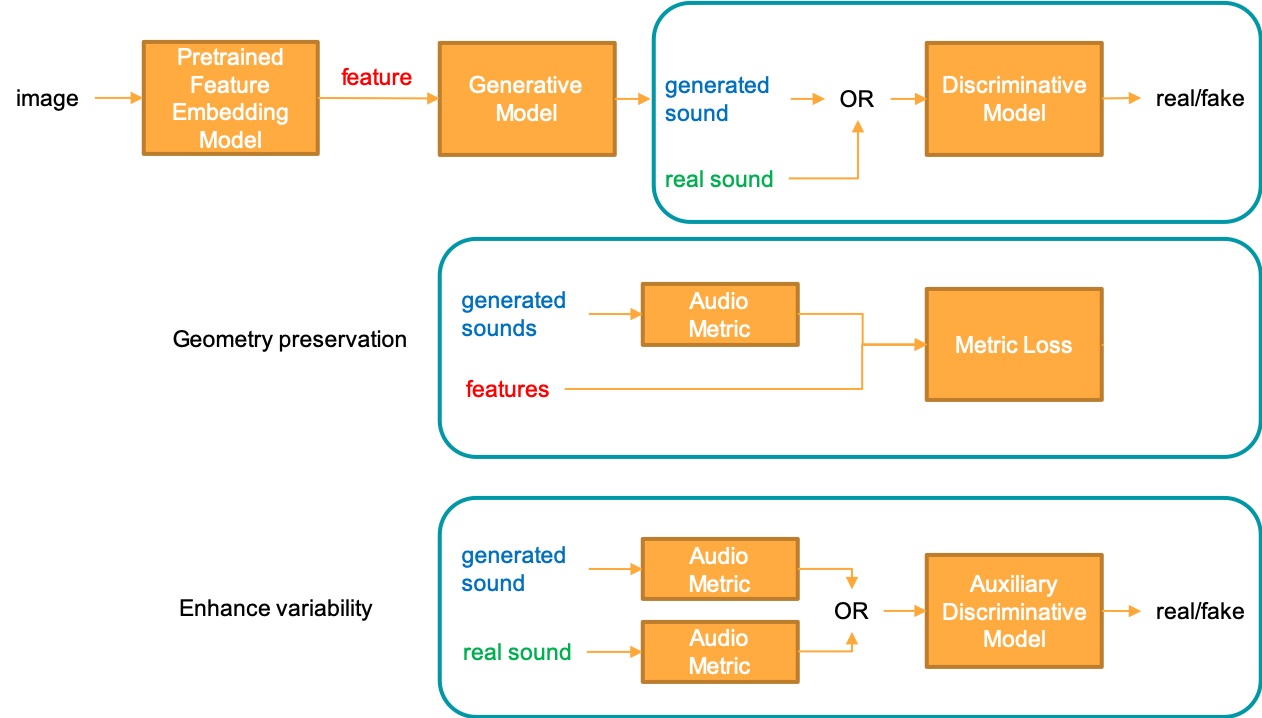}
	\caption{Our proposed image-to-audio translation model's structure. We use ``OR" here to indicate that the input batch comes from (exclusively) one of the two inputs. 
	}
	\label{fig:network}
\end{figure}

Our method employs the distance-preserving mapping technique~\cite{port2020earballs, onesided} in the GANs framework: We first find a feature embedding model that embeds input face images into a metric space, where the distance corresponds to a measure of face similarity. This embedding model can be obtained by refining a convolutional neural network that is pretrained for the face recognition task on a large-scale face dataset (VGGFace2~\cite{Cao18}), where the network captures information about the facial appearance to successfully classify the face identities. We then train a deep generative model by using a GAN approach~\cite{goodfellow2014generative} that takes the face features from that embedding space~\footnote{The general GANs take input features from a simple latent distribution-such as the Gaussian or uniform distribution. For our task, we are using the pretrained feature embedding space as the input latent.} and synthesizes raw audio waveforms that fit into the distribution of sounds specified by a given target dataset (i.e. musical notes of NSynth dataset~\cite{nsynth2017}). A discriminator is simultaneously trained with the generator, whose task is to predict whether the generated output is real or fake. See the overview of our model's structure in Fig.~\ref{fig:network} Top row. 

To enforce the distance preservation constraint, we add a metric loss term (Eq.~\ref{eq:metric-loss}) to the GAN's adversarial loss (Eq.~\ref{eq:adv_cost_func_G}), which computes the difference of pairwise distances of the face features and features of the generated audio samples. See Fig.~\ref{fig:network} Middle row. 

The audio feature is computed by a combination of a pretrained audio feature embedding model and a MFCCs feature (The 'Audio Metric' block in Fig.~\ref{fig:network}): We first train a feature embedding model for audio that maps raw audio waveforms into a compact Euclidean space where the distance directly corresponds to a measure of audio similarity. This is done by using the triplet loss function~\cite{schroff2015facenet} that aims to separate the positive pair from the negative sample by a distance margin, where the positive and negative are determined by the annotated instrument family class label of the NSynth dataset. As a result, audio samples that have the same musical instrument timber (e.g. violin, keyboard, etc) are mapped into features with small distances and audio samples that have distinct musical instrument timbers are mapped to features with larger distances. To further incorporate the perceptual distance measure into the audio metric, the MFCCs feature of the audio sample is computed and concatenated with the learned audio features. See more details in Section~\ref{sec:audio_metric}.

Lastly, to enhance the fidelity of the synthesized sounds, an additional discriminator is attached on top of the audio features (i.e. audio samples passed the audio metric block as in Fig.~\ref{fig:network} Bottom), whose task is to distinguish between features computed from real audio and the generated audio.

In the following sections, we will describe more details on each component. 

\subsection{Face Feature Embedding Model}
Our face feature embedding model is a convolutional neural network that is trained on 
MS-Celeb-1M dataset~\cite{guo2016ms} and then fine-tuned on VGGFace2 dataset~\cite{Cao18} for the face recognition task~\footnote{The pre-trained model and weights are available from the VGGFace2~\cite{Cao18} authors' project page.}. 
This embedding network follows ResNet-50 architecture ~\cite{he2016deep} and we take the output of the second last fully connected layer of this network (i.e. before the top classification layer) and apply L2 normalization. Overall, the model takes 224x224 face images as input and outputs 512-dimensional unit vectors as the features. 
This feature embedding captures the high-level information about the facial appearances of the training images and have a Euclidean space where a measure of face similarity corresponds to the squared L2 distance. i.e. faces of the similar appearance have small distances and visually distinct faces have large distances.   



\subsection{Adversarial Learning}
To enforce that the output sounds fit into a target distribution (i.e. 
musical sounds~\cite{nsynth2017}), we use a discriminator network (Fig.~\ref{fig:network} Top row).  This network is tasked with predicting whether the sounds fed into it are ``real" audio samples from the target dataset or synthetic audio samples output by the generator, and we trained this network simultaneously with the generator by the min-max adversarial learning approach~\cite{goodfellow2014generative}.  

 
In the adversarial training, the generator G and the discriminator D are trained by the following standard objective function of GAN~\cite{goodfellow2014generative}:
\begin{equation}\label{eq:adv_object}
\min_{G}\max_{D} V(G,D)=\min_{G}\max_{D} \mathop{E}_{y \sim q_{data}}[\log{D(y)}] + \mathop{E}_{z \sim p_{data}}[\log{1-D(G(z))}]
\end{equation}
where $z$ is the pre-trained face feature embedding model's output of the input face image that is sampled from a source distribution $p_{data}$ (i.e. VGGFace2 dataset), and $y$ is the real audio data sampled from a target distribution $q_{data}$ (i.e. NSynth dataset). For the updates of D and G, we use the following cost functions as in~\cite{miyato2018spectral} respectively:
\begin{equation}\label{eq:adv_cost_func_D}
V_D(\hat{G}, D) = 
\mathop{E}_{y \sim q_{data}}[ \min(0, -1 + D(y)) ] +
\mathop{E}_{z \sim p_{data}}[ \min(0, -1 - D(\hat{G}(z)) ]
\end{equation}
\begin{equation}\label{eq:adv_cost_func_G}
V_G(G,\hat{D}) = -\mathop{E}_{z \sim p_{data}}[\hat{D}(G(z))]
\end{equation}

So far this training scheme is similar to that described in~\cite{wavegan}, however, we differ in the following ways.  To stabilize the adversarial min-max training between two networks, we employ the spectral normalization technique from~\cite{miyato2018spectral} which enforces a Lipschitz constraint on the discriminator.  Additionally, in order to incorporate conditional information while training the generator and the discriminator network (e.g. pitch class label in NSynth~\cite{nsynth2017} musical sounds dataset), the conditional batch normalization~\cite{ghiasi2017exploring,perez2018film} and the projection-based method by ~\cite{miyato2018cgans} are used, respectively. Hence, our generator and discriminator are based on the models proposed in ~\cite{wavegan} except that the weight are normalized with the spectral normalization technique and the additional conditional projection layers are added. 

\subsection{Metric Preservation Constraint}\label{sec:metric-loss}

To enforce that our translation from image to sound is preserving the geometry of the image domain metric space, we follow the implementation studied in~\cite{onesided}, which uses the mean absolute difference between the standardized pairwise distances in the two metric spaces.  In specific, we define our metric preservation loss as 
\begin{equation}\label{eq:metric-loss}
\mathcal{L}_{metric} = \displaystyle \frac{1}{Z}\sum_{i<j} {
\left| \frac{||f(x_i)-f(x_j)||_2 - \mu_x}{\sigma_x} - \frac{||\phi(y_i)-\phi(y_j)||_2 - \mu_y}{\sigma_y} \right|
}
\end{equation}
where $N$ is our batch size, $Z$ is the number of possible (unordered) pairs of $N$ samples
,  $f$ is the fixed feature embedding model, $\phi(y)$ represents our audio metric of the translated audio output $y$. The standardization parameters, ($\mu_x$, $\sigma_x$) and ($\mu_y$, $\sigma_y$), are derived from source and target datasets respectively and are the mean / standard deviation of pairwise distances of samples from within each respective dataset.  


In the update of the generative network in the adversarial min-max training, this metric loss is added to the adversarial loss Eq.~\ref{eq:adv_cost_func_G}:
\begin{equation}\label{eq:adv_cost_func_G_metric}
V_G(G,\hat{D}) = -\mathop{E}_{z \sim p_{data}}[\hat{D}(G(z))] + \lambda \mathcal{L}_{metric}
\end{equation}
where $\lambda$ controls the balance between the adversarial and metric losses. We set it to 10 throughout our experiments.



\subsection{Audio Metric}\label{sec:audio_metric}
Our audio metric is designed to take two types of information into account: the Mel-frequency cepstral coefficients (MFCCs)~\cite{MelScale} and also any useful extra class labels/metadata available the modeller wishes to incorporate. For example, in the NSynth dataset~\cite{nsynth2017}, each audio sample is tagged with additional metadata such as instrument family (e.g. guitar, flute, etc.), pitch, human-labeled quality, and so on. In order to take this discrete metadata into account we use a convolutional neural network to learn an embedding from the target dataset of audio samples into a Euclidean space, where audio samples of the same class have small distance and audio samples of different class have large distance. We then concatenate this learned embedding with the MFCC transform (Eq.~\ref{eq:audio_metric}) to give us a fixed embedding for audio samples.  We use the L2 norm on top of this embedding for our audio metric.

We sacrifice the end-to-end trainability of our system to learn this dataset-specific audio metric, however, for our musical target space, we believe the extra information helped our generator learn to produce sounds which were easier for humans to discern between and whose differences were more memorable.

In specific, our audio embedding is given by 
\begin{equation}\label{eq:audio_metric}
\mathbf{\phi} = \left[  
  \lambda_1 \mathbf{\psi} ( y ) ^ T, \lambda_2 MFCC(y)^T
\right]^T
\end{equation}
where $y$ is the input audio signal and $\psi(\cdot)$ represents the trained embedding network and $MFCC(\cdot)$ denotes the array of MFCCs flattened to one-dimensional vector. Here $\lambda_1$ and $\lambda_2$ are precomputed normalization factors set to the the maximum distance between any two audio samples in the target dataset according to $\psi(\cdot)$ and $MFCC(\cdot)$ respectively.

The audio embedding network is trained using triplet loss (\cite{schroff2015facenet}) and outputs a 128-D unit vector. For the MFCCs, we used an implementation suggested in the TensorFlow (\cite{abadi2016tensorflow}) documentation and flattened the output to a 377-D feature vector. The concatenated feature $\mathbf{f}$ becomes a 505-D vector. 

Fig.~\ref{fig:viz} (b) shows our customized audio metric space of NSyth samples. For the visualization, we mapped them into a 2D space by tSNE~\cite{maaten2008visualizing}. Each blob represents an NSynth sample's feature vector as given by Eq.~\ref{eq:audio_metric} and is color-coded by the instrument family. Notice that audio samples in this audio metric are grouped by the incorporated labeling information i.e. the 10 instrument family: (bass, brass, flute, guitar, keyboard, mallet, organ, reed, string, and vocal). Each red dot represents the audio translation of a face from VGGFace2. 


\subsection{Enhancement of the Sample Variety}\label{sec:enhancement}
The adversarial loss term in Eq.~\ref{eq:adv_cost_func_G_metric} enforces that the generated audio fits into the given target audio dataset. Despite this term, when using a metric preservation constraint (i.e. the second term in Eq.~\ref{eq:adv_cost_func_G_metric}), we observe decreased variety in the generated audio.  We verify this perceived lack of variety using Fr\^echet Inception Distance (FID)~\cite{heusel2017gans}. Our FID score results and a visual check can be seen in Table~\ref{tab:performance} and Figure~\ref{fig:viz} respectively. 


One might notice that $\lambda$ in Eq.~\ref{eq:adv_cost_func_G_metric} can be adjusted to balance out between the variety and the metric preservation, but it is a time consuming process to find the optimal value of $\lambda$. Instead, we set it to a fixed value that guarantees the metric preservation (e.g. 10.0) and we add au auxiliary discriminator as an additional adversarial criterion that can support the main discriminator in a way that it further encourages the generative model to output audio samples that fit to the distribution of the target audio dataset. Note that, unlike the main discriminator, we attach this second discriminator on top of our audio metric feature (Fig.~\ref{fig:network} bottom row) to predict the real/fake by processing the audio features rather than the raw audio signals. As we will see in the next section, this helps the generative model output audio samples that are spread over the distribution of real audio samples in our audio metric space (Fig.~\ref{fig:viz} (b) third row). Furthermore, our additional discriminator has a much simpler and economical architecture than the main discriminator, which consists of a series of a fully connected layer-nonlinear activation layer blocks. The detailed network structure of the second discriminator will be described in \appendixsection.

Our final update functions for the generative and the two discriminative networks are as follows.
\begin{equation}\label{eq:adv_cost_func_D_final}
\begin{split}
V_{D_{1,2}}(\hat{G}, D_1, D_2) = & \mathop{E}_{y \sim q_{data}}[ \min(0, -1 + D_1(y)) ] + \\
& \mathop{E}_{z \sim p_{data}}[ \min(0, -1 - D_1(\hat{G}(z)) ] +\\
& \mathop{E}_{y \sim q_{data}}[ \min(0, -1 + D_2(\phi(y))) ] +\\
& \mathop{E}_{z \sim p_{data}}[ \min(0, -1 - D_2(\phi(\hat{G}(z))) ]
\end{split}
\end{equation}
\begin{equation}\label{eq:adv_cost_func_G_final}
\begin{split}
V_G(G,\hat{D}_1, \hat{D}_2) = & -\mathop{E}_{z \sim p_{data}}[\hat{D}_1(G(z))] \\
& -\mathop{E}_{z \sim p_{data}}[\hat{D}_2(\phi(G(z)))] + \lambda \mathcal{L}_{metric}
\end{split}
\end{equation}
where $D_1$ and $D_2$ are the primary and the auxiliary discriminative networks and $\phi(\cdot)$ is the audio metric function defined in Section~\ref{sec:audio_metric}.


\section{Results}
In this section, we first describe our evaluation metrics used for assessing our translation model. We then compare our translation model to: 1) a baseline model, 2) the same model but with the distance metric preservation , 3) our final model that uses the auxiliary discriminator to enhance the variety of the generated audio, and 4) an adversarial autoencoder type of model by~\cite{makhzani2015adversarial}. Lastly, we will preset our experimental results. 

\subsection{Evaluation Metrics}
We use four metrics in total. Pearson product momentum correlation (PC) and silhouette value (SV) are used to measure the metric preservation.  To score the quality of the translated audio quality and how well it fits into the target dataset, we use two evaluation metrics specially designed for GANs: Inception Score (IS)~\cite{salimans2016improved} and Fr\^echet Inception Distance (FID)~\cite{heusel2017gans}. 

\subsubsection{Inception Score (IS)}
The IS uses an image classification model built on top of inception base~\cite{szegedy2015going} and rewards a generative model that outputs a sample which is easily classified into a single class by the inception classifier, as well as a model that outputs samples belong to all the possible classes. The drawback of IS is that it does not measure the quality of the generated samples compared to the real samples and does not penalize the lack of diversity in the generated samples~\cite{heusel2017gans}. Higher IS indicates higher quality audio samples.

\subsubsection{Fr\^echet Inception Distance (FID)}
FID is based on the 2-Wasserstein distance (also called Fr\^echet distance) between Gaussians fit to the inception feature vectors for the real and the generated samples. It measures how similar two groups of samples are in terms of the statistical mean and variance of the feature vectors. FID is known to be sensitive to both the fidelity and variety of the generated samples~\cite{brock2018large}. 

To the best of the authors' knowledge, unlike the image domain, there is no standard and public inception structure based classification model for the audio domain. Thus, for FID and IS, we trained our own classification network that estimates the pitch (or the instrument family) label on the NSynth training samples and used the outputs of its second last layer (i.e. before the last classification layer) to compute the scores. To obtain the FID score of the generated samples, we compare the statistics of our pretrained model's output feature maps from the generated samples with those from NSynth's training samples. Lower FID indicated higher quality of audio samples.

\subsubsection{Pearson product momentum correlation (PC)}
Our next evaluation metric is for measuring how well the geometry of the source metric space is preserved in the target audio metric space. To do this, we compute the Pearson product momentum correlation (PC) between L2 pairwise distances of the samples in the source metric and the corresponding translated samples in the audio metric. The Pearson correlation has a value between +1 and −1, where 1 is total positive linear correlation.

\subsubsection{Clustering evaluation by Silhouette Value (SV)}\label{sec:sv}
As our target audio metric was designed to take class label information into account as in Section~\ref{sec:audio_metric}, and thus the audio samples are grouped by the class labels in the audio metric space (e.g. see 10 clusters of NSynth samples in Fig.~\ref{fig:viz} (b)), we design a metric that inversely checks whether the clusters are preserved in the source metric space. More precisely, we first assign the class label to each translated audio clip based on its closest cluster of ``real" audio samples in the audio metric space. I.e. a translated audio sample which lies within or close to the `string' cluster will be assigned the `string' family label. Then, we measure how well the corresponding untranslated samples are grouped by their assigned class labels in the source metric space. To measure this quantitatively, we use Silhouette value which is presented by ~\cite{rousseeuw1987silhouettes} and has been used as a means for clustering evaluation~\cite{kim2019info}. The classified face samples in the source metric space are visualized in Fig.~\ref{fig:viz} (a). 

\subsection{Models}
\subsubsection{Baseline}
Our baseline model is based on WaveGAN~\cite{wavegan}. It is trained to generate from an input 512-dim VGGFace2 feature vector a raw waveform of audio signal of length 8192 samples (around 0.5 sec by 16000 Hz sample rate)~\footnote{We use only first slice from each NSynth audio.} that fits to a distribution of NSynth training samples. To get better quality of sounds, we employed a state-of-the-art stabilization technique, spectral normalization technique~\cite{miyato2018spectral}. More precisely, all the weight matrices in the deconvolutional and fully connected layers of WaveGAN model are normalized by their estimated first singular value. Further, our baseline model is conditioned by a 'pitch' label given by the NSynth dataset (i.e. our model is the \textit{conditional} GAN similar to ~\cite{engel2019gansynth}) by adding conditional batch normalization layers~\cite{ghiasi2017exploring,perez2018film} in the generative network and adding a projection layer at the end of the discriminator as in ~\cite{miyato2018cgans}. Note that, through our experiments, we generated audio signals from the face feature vectors with random pitch values.  

\subsubsection{Proposed model}  
We performed an ablation study on our proposed model by adding sequentially the metric preservation loss and then the second discriminator to the baseline to see the effect of each constraint. 

\subsubsection{Adversarial AutoEncoder model}~\cite{makhzani2015adversarial}
We trained an adversarial autoencoder model proposed by ~\cite{makhzani2015adversarial} on the NSynth audio samples and modified it to enforce that the encoder outputs latent vectors that fit to the distribution given by VGGFace2 face feature vectors. Once this network is trained, the decoder can be used as a generative model that maps the face embeddings to the NSynth audio signals. To get better quality of the audio sounds, we further enforce that the output audio of the decoder as close to real audio as possible by an additional discriminator (like our baseline model) and its loss term is added to the original reconstruction loss term. 

\subsection{Experimental results}
\begin{figure}[t!]
\centerline{
\begin{tabular}{cc}
 \includegraphics[width=\x]{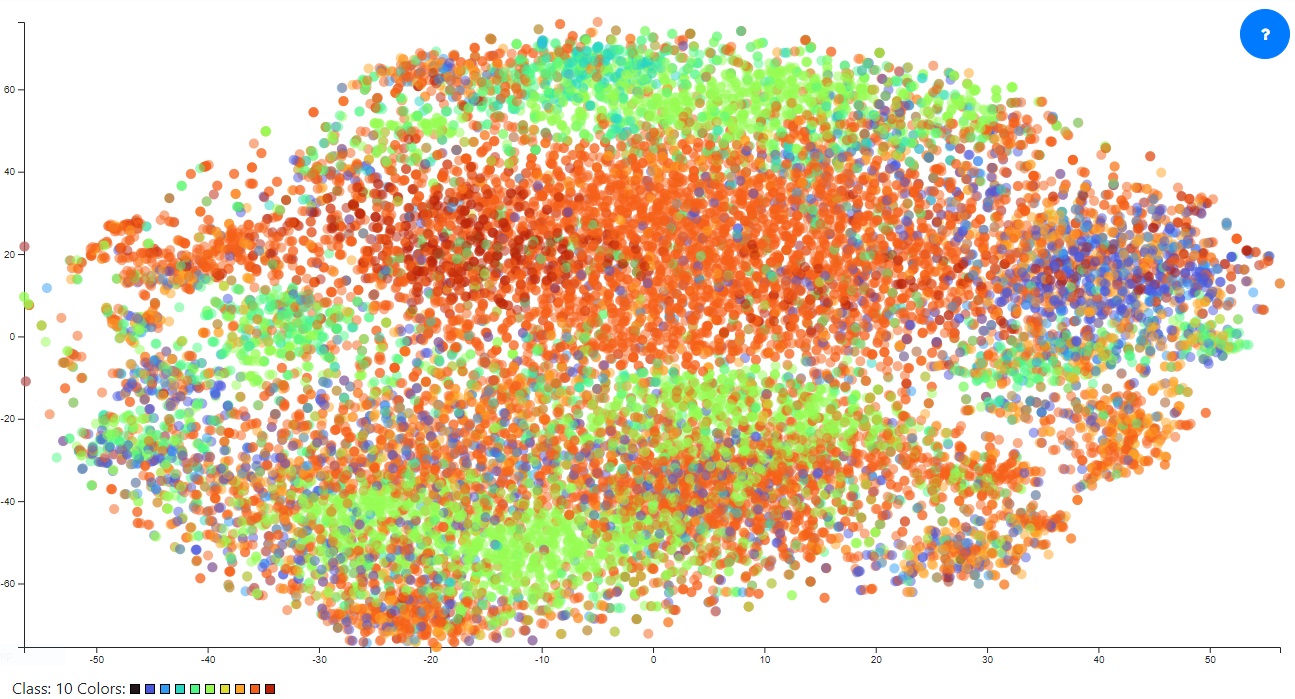} &
 \includegraphics[width=\x]{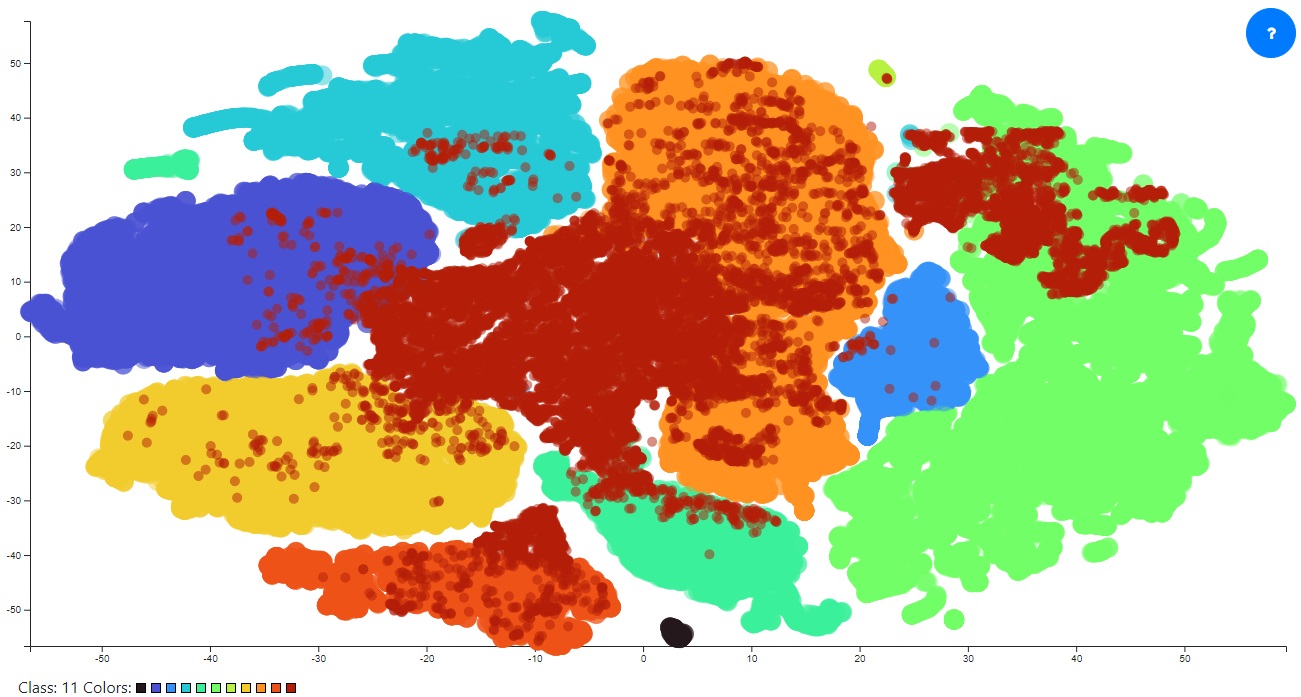} \\
 \includegraphics[width=\x]{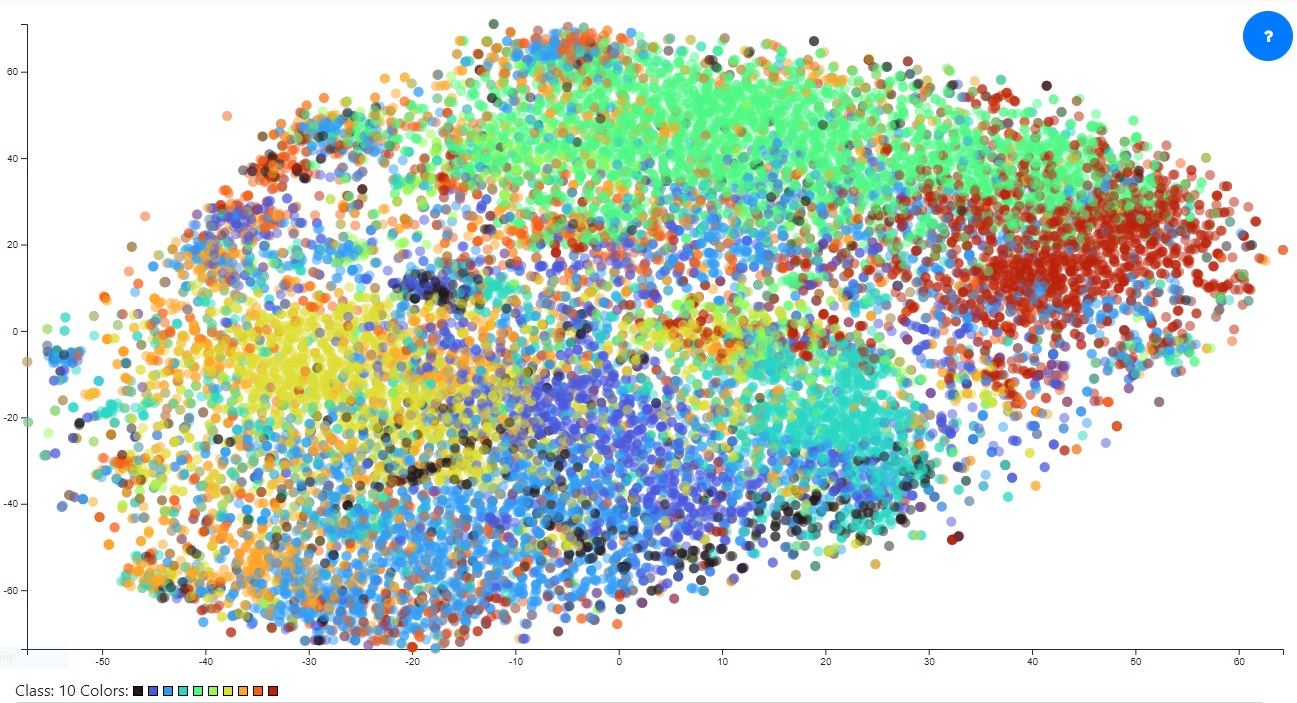} &
 \includegraphics[width=\x]{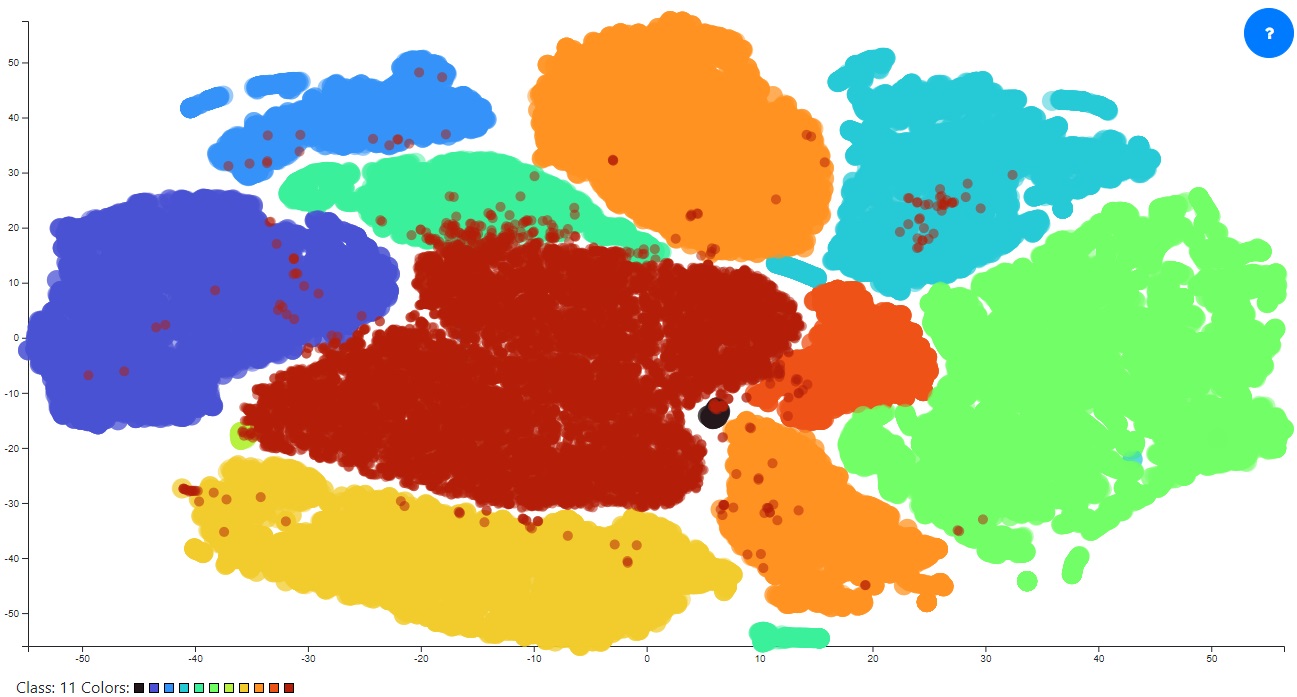}\\
 \includegraphics[width=\x]{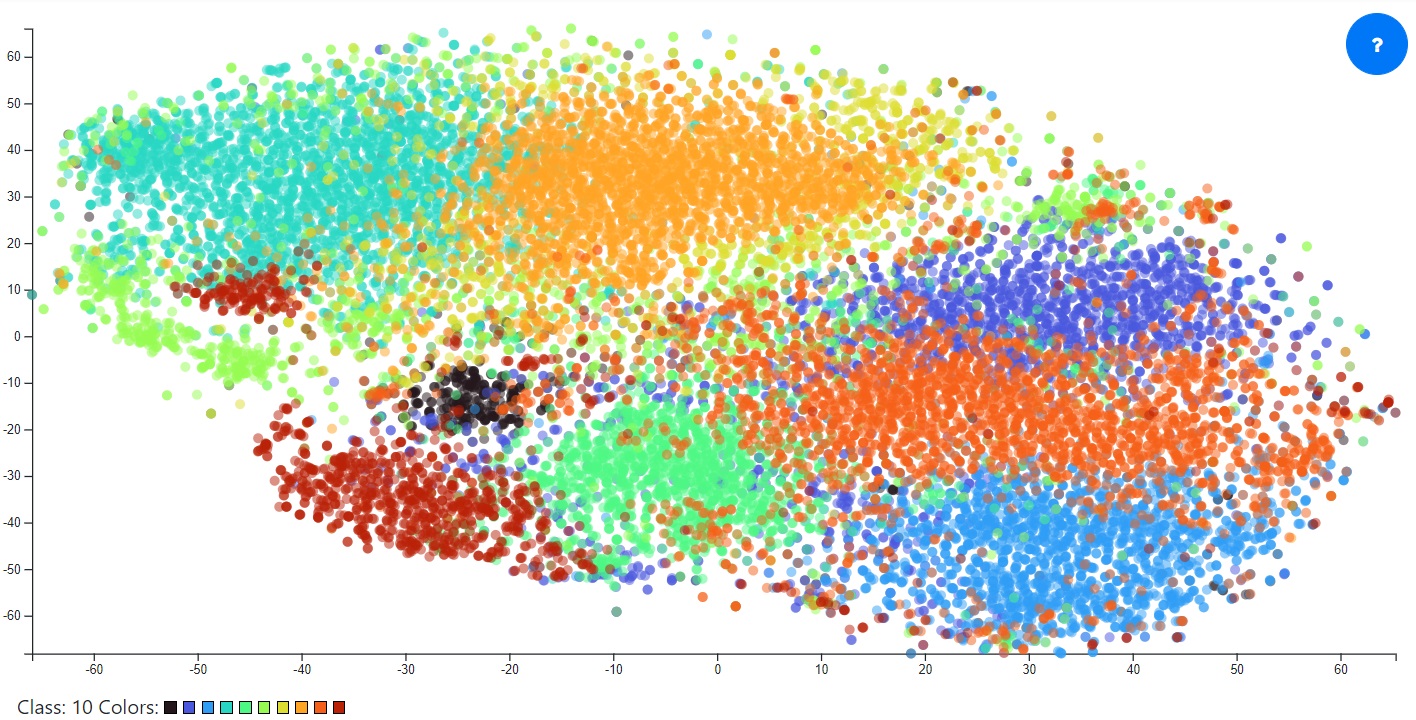} & 
 \includegraphics[width=\x]{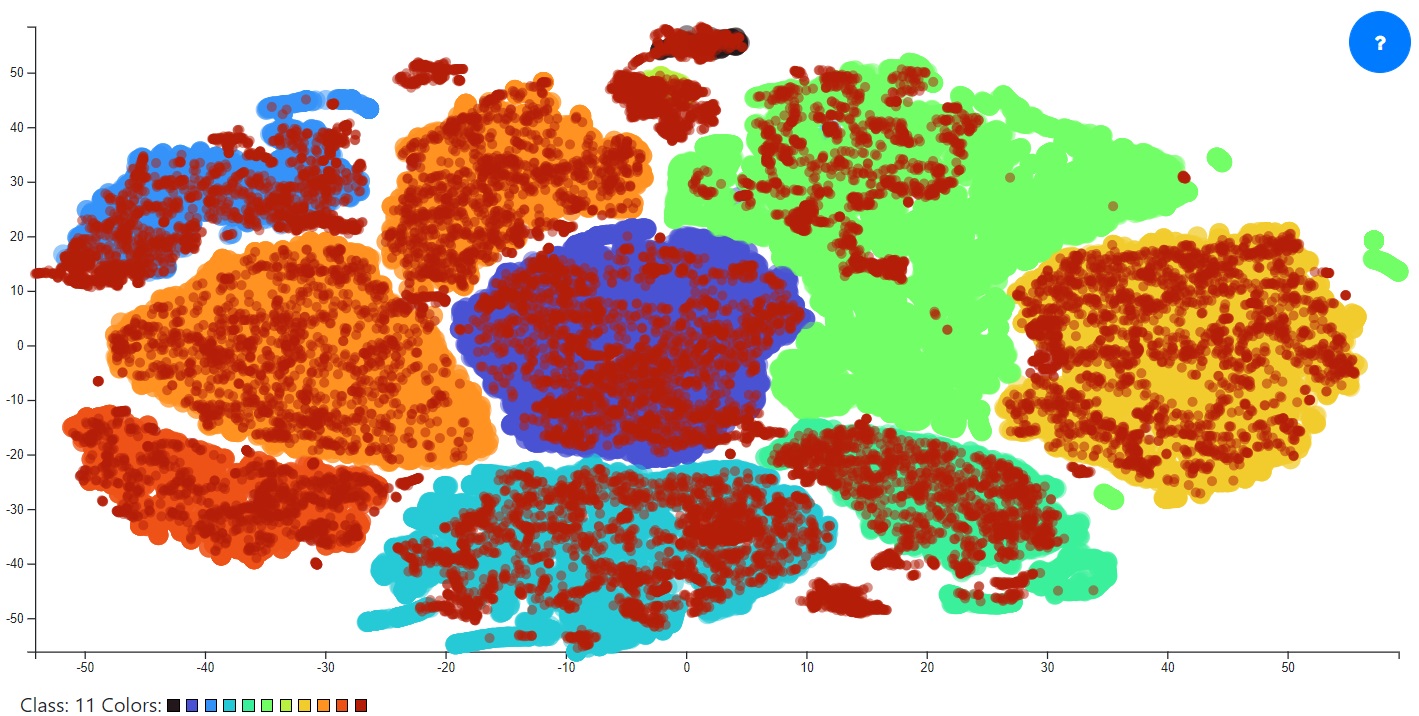} \\
 \includegraphics[width=\x]{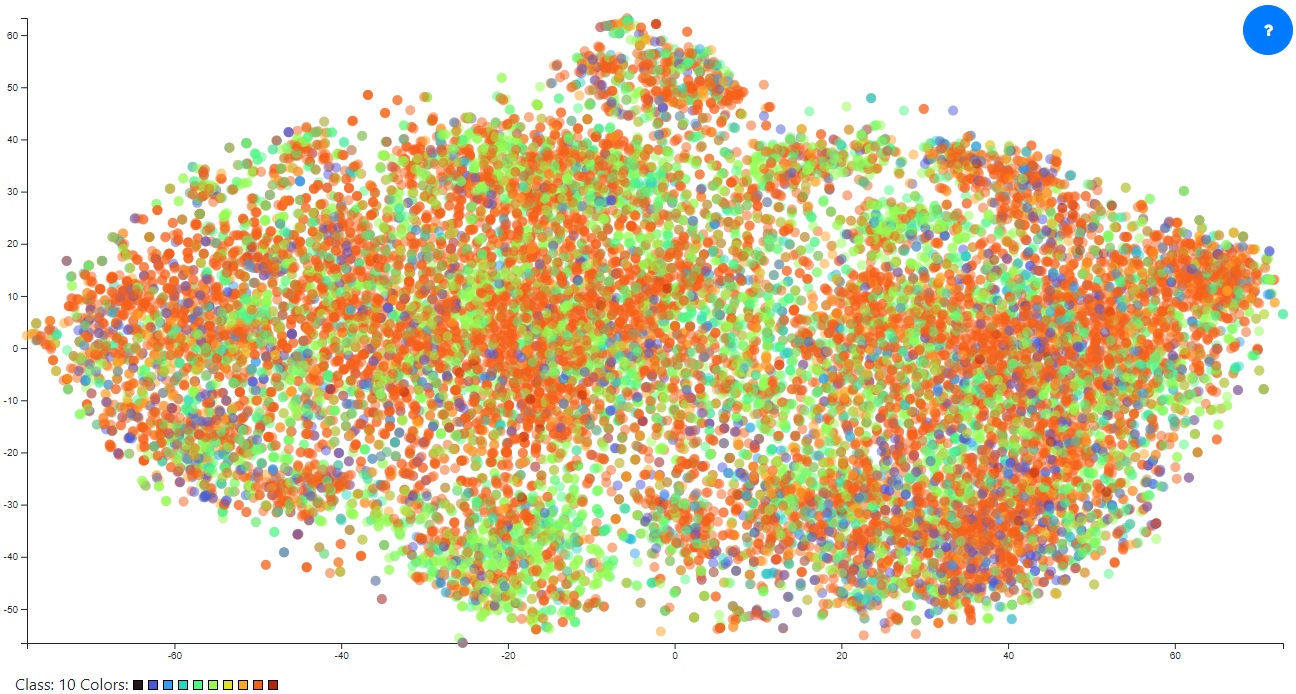} & 
 \includegraphics[width=\x]{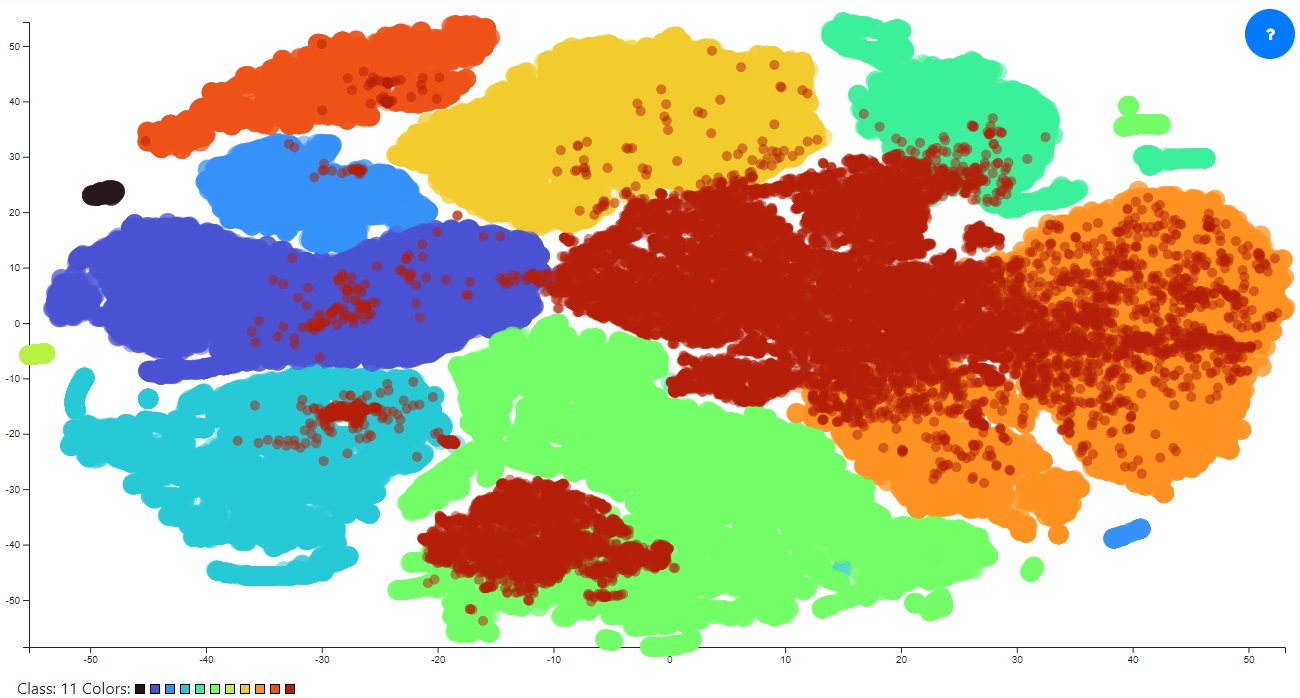}\\
 (a) Face metric space & (b) Audio metric space
 \end{tabular}
}
\caption{(a) tSNE visualization of the source face embedding metric space with 20k VGGFace2 face samples. Each face sample is color-coded by its estimated instrument family label (See text for details). (b) tSNE visualization of the target audio metric space with the NSynth real audio samples (color-coded blobs) and translated face samples (red dots). From Top to Bottom: 1) baseline model 2) w/ distance preservation 3) w/ auxiliary discriminator 4) conditional adversarial autoencoder~\cite{makhzani2015adversarial}. The colors represent the 10 instrument families: bass, brass, flute, guitar, keyboard, mallet, organ, reed, string, and vocal. A video demo of our proposed model’s learned translation (i.e. third row in (a)) is available in \videodemoaddress.
}
\label{fig:viz}
\end{figure}

Table~\ref{tab:performance} shows the above four evaluation metric scores. First, the baseline model achieves the FID and IS scores of 21.84 (51.02) and 56.27 (4.59) respectively by our pre-trained pitch (family) classification model~\footnote{Note that the maximum of IS score is the number of classes which is 61 for the pitch class and 10 for the family class. The minimum score of FID is zero for the both classification models.}. As we use our own classification models, as a reference, we also report the FID score computed on the NSynth's test samples~\footnote{We use the same train/test split as in~\cite{engel2019gansynth}.}, which is 1.25 (0.72) by the pitch (family) classification model. The IS score computed on the NSynth's test set is 56.73 (6.78). 
With the distance metric preservation constraint on the model, the Pearson Correlation (PC) score increases from 0.090 to 0.683. This demonstrates that our metric preservation loss helps the model preserve the geometric structure of the face embeddings in the target audio metric space, however, we also observed a reduced quality in the samples, i.e FID increases from 21.84 (51.02) to 46.69 (90.00), and IS decreases from 56.27 (4.59) to 53.18 (3.98). 
Our visualization of the translated audio samples in the audio metric space also demonstrates this. See Fig.~\ref{fig:viz} (b): the red dots represent the translated audio clips from randomly sampled 20k faces in the VGGFace2 train set, and blobs are the NSynth real audio samples. 
The first two top figures correspond to the baseline and the one with metric preservation loss. It is apparent that the distribution of red dots (i.e. translated audio samples) shrinks and forms a cluster surrounded by other ``real" clusters and we find that most of those generated audio samples play interpolated sounds between the nearby instrument families (clusters) and hard to find samples that play other instrument families at a distance from them (e.g. the right most green cluster (`mallet') and left top light-blue cluster (`flute')). 

By adding an additional adversarial constraint by the auxiliary discriminative network as in Section~\ref{sec:enhancement}, we improve the variety while it still preserves the source metric: 25.52 (FID) and 0.432 (PC). We also observed that the auxiliary discriminator changes a lot the translations' distribution in the audio metric space in a way that they are spread evenly over the real clusters (See the third plot in Fig.~\ref{fig:viz} (b)) and the translations play much wider variety of musical sounds.

We also measured that how well faces are grouped by their assigned labels in the source metric space by the Silhouette value (See Section~\ref{sec:sv}) and our model with the auxiliary discriminator outperforms others by 0.0256. Note that overall SV scores are low since the pre-trained face embedding was not designed to have such well separated clusters, but the SV rewards the result where faces with the same class sit close together in the space.  Fig.~\ref{fig:viz} (a) shows the same randomly sampled 20k face features in the source metric space color-coded by the assigned labels. Our translation model with the auxiliary discriminator (the third one) clearly shows the clusters by the assigned labels. 

Lastly, we find our GAN based model outperforms the autoencoder type of model~\cite{makhzani2015adversarial} in this audio generation task in terms of both the quality (FID, IS scores) and the preservation (PC). Unlike the visualization in ~\cite{makhzani2015adversarial} which shows the smoothed variation in the generated images with the one in the latent code, we didn't find such a smoothed mapping which would result in the similar faces are mapped to similar musical sounds.

\begin{figure}[t!]
\centerline{
\begin{tabular}{c}
 \includegraphics[width=\y]{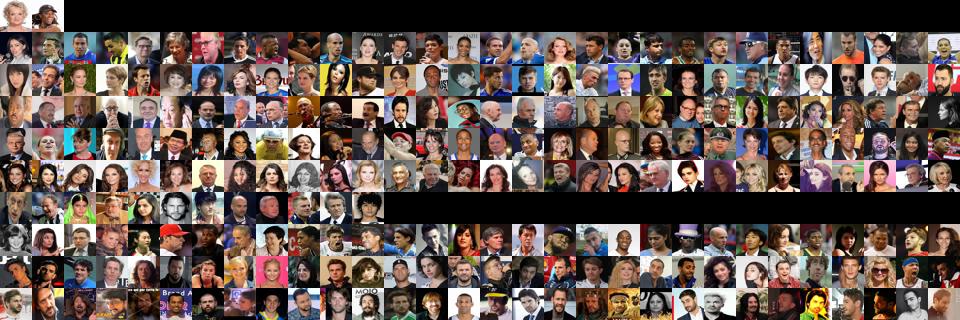} \\
 (a) Baseline\\
 \includegraphics[width=\y]{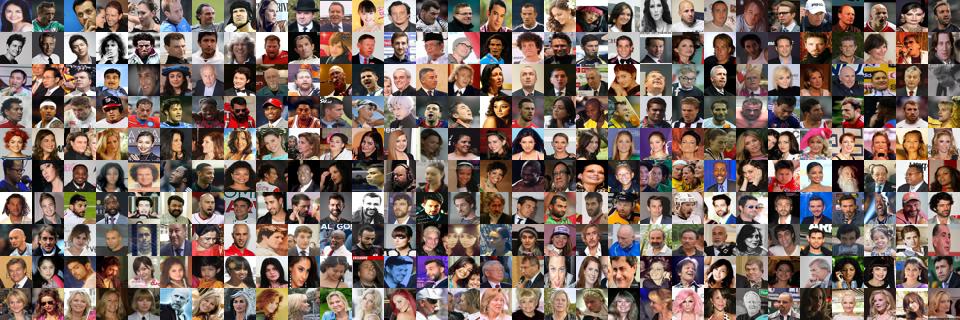} \\
 (b) Baseline + metric preservation \\
 \includegraphics[width=\y]{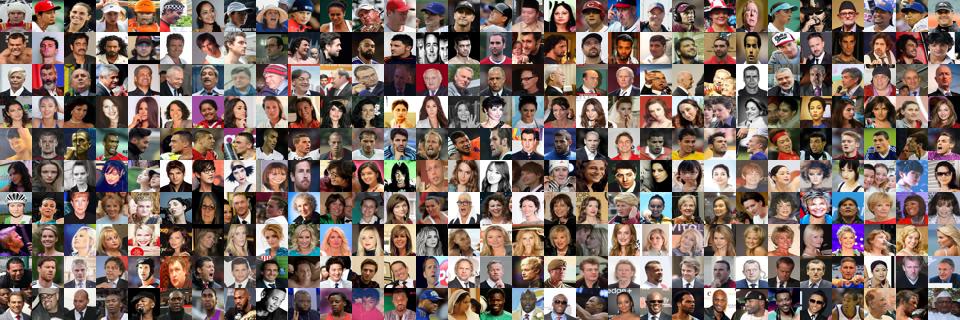} \\
 (c) Baseline + auxiliary discriminator \\
 \includegraphics[width=\y]{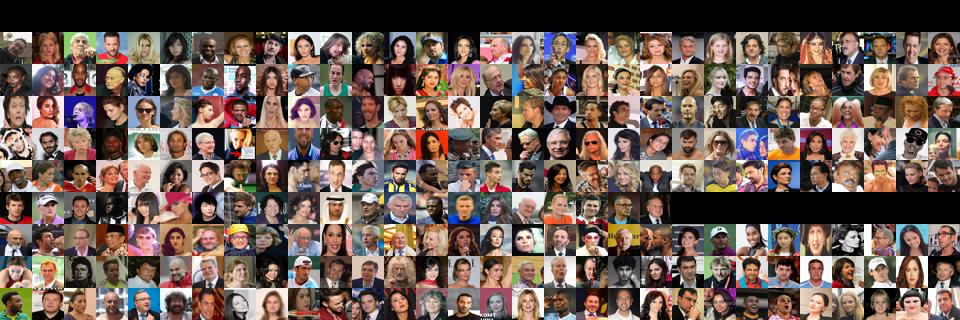} \\
 (d) Adversarial Autoencoder~\cite{makhzani2015adversarial}
 \end{tabular}
}
\caption{
Randomly sampled face images groped by their instrument class labels. Fig.~\ref{fig:viz} (a)) to visually check if those faces show a pattern on their visual appearance (e.g. same skin tone, hair color, etc.). Our translation model (c) shows faces belonging to the same instrument family are visually similar to each other and show one or more common attributes (e.g. female/blond hair in the seventh row). If the number of faces per class is fewer than 30, that row is padded by black pixels.
}
\label{fig:sprite_face}
\end{figure}

We also randomly sampled face images belonging to the instrument family label to visually check if those faces show a pattern on their visual appearance (e.g. same skin tone, hair color etc.). See Fig.~\ref{fig:sprite_face}: each row shows face images belonging to the same instrument family (i.e. from the dots with the same color in Fig.~\ref{fig:viz} (a)). Our translation model (Fig.~\ref{fig:sprite_face} (c)) shows faces belonging to the same instrument family are visually similar to each other and show one or more common attributes (e.g. female/blond hair in the eighth row).

\begin{table}[t]
\centering
\caption{The quality measure scores of Fr\^echet Inception Distance (FID, lower is better) and Inception Score (IS, higher is better) by a pitch classifier and a family classifier, the distance preservation measure score by Pearson Correlation (PC), and the clustering measure score by Silhouette Value (SV). 
\vspace{-3mm}
\label{tab:performance}}
\resizebox{3.0in}{!}{%
{\renewcommand{\arraystretch}{1.4}%
\begin{tabular}{ l | c | c | c | c | c | c }
     \hline
     & \multicolumn{2}{c|}{Pitch} & \multicolumn{2}{c|}{Family} & &\\
     \hline
     \textbf{Method} & \textbf{FID} & \textbf{IS} & \textbf{FID} &  \textbf{IS} &  \textbf{PC} & \textbf{SV} \\
     \hline
    \textbf{Baseline} & 21.84 & 56.27 & 51.02 & 4.59 & 0.090 & -0.0051\\
    \textbf{+goe-preservation} & 46.69 & 53.18 & 90.99 & 3.98 & 0.683 & 0.0112\\
    \textbf{+aux. discriminator} & 25.52 & 54.97 & 53.98 & 5.89 & 0.432 & 0.0256 \\
    \hline
    \textbf{AAE~\cite{makhzani2015adversarial}} & 53.06 & 49.43 & 94.07 & 4.18 & 0.142 & -0.0036\\
    \hline
  \end{tabular}
  }
  }
\end{table}

\section{Conclusions}
We have proposed a distance preserved generative adversarial network that automatically learns an information preserving embedding between two unrelated domains of media (i.e. image and audio domain). We discovered that there is a trade-off between the variety of translations and the preservation of geometric information. To address this problem, we proposed to use the auxiliary discriminator that can support the primary discriminator, which shows the balanced performance between the diversity and the metric preservation. We demonstrate that the proposed model translates a face image from the VGGFace2 dataset into a musical sound that plays one of 10 instrument family in the NSynth dataset and the faces playing the same instrument type show a pattern on their visual appearance (e.g skin tone, hair color etc.).

%% file: appendix.tex
In this section, we describe details about implementation of our deep neural network models.

\subsection{Network Architectures}
\subsubsection{Face Embedding Network}
The face embedding network is based on a pretrained ResNet-50 architecture ~\cite{he2016deep} and we take the output of the second last fully connected layer of the network (i.e. before the top classification layer) and apply the average pooling and L2 normalization to the output. Table~\ref{table:face_embedding} shows detailed configuration of the network architecture.

\begin{table} [ht]
\centering
\caption{Our face embedding network's structure. In the parameters column, d is the number of outputs.}
\begin{tabular}{ lll } 
 \hline
 Operation & Parameters & Outputs \\ 
 \hline
 Input image & & 224 x 224 x 3\\
 ResNet-50 & & 7 x 7 x 2048 \\
 AveragePooling2D & & 1 x 1 x 2048 \\
 Flatten & & 2048 \\
 Dense & 512d & 512 \\
 L2-normalize & & 512 \\
 \hline
\end{tabular}
\label{table:face_embedding}
\end{table}

\subsubsection{Generator and Discriminator Networks}
The generator network is based on the WaveGan structure~\cite{wavegan} that uses the 1-D transposed convolution operation to upsample low-resolution feature maps into a high-resolution audio signal. See Table~\ref{table:generator}. To incorporate the conditional information to the network, all the the batch normalization layers are conditioned by the class labels (i.e. pitch labels) as in ~\cite{ghiasi2017exploring,perez2018film}.

The discriminator network structure is also similar to the WaveGan except that we conditioned the network by the pitch label using the projection method as in~\cite{miyato2018cgans} and the spectral normalization was applied to all weight matrices in the layers to enforce a Lipschitz constraint. See Table~\ref{table:discriminator}.  

\begin{table} [ht]
\centering
\caption{Our generator network's structure. The UpConv1DBlock network consists of Conv1DTranspose - Conditional Batch Norm~\cite{ghiasi2017exploring,perez2018film} - LeakyRELU layers. The last UpConv1DBlock uses tanh activation layer instead of the LeakyRELU. In the parameters column, k is the kernel length, s is the stride, d is the number of outputs.}
\begin{tabular}{ lll } 
 \hline
 Operation & Parameters & Outputs \\ 
 \hline
 Input(face feature vector, pitch label) & & 512, 1\\
 Dense & 4*4*64*32d & 4*4*64*32\\
 Reshape & & 16 x (64*32) \\
 Conditional Batch Norm & & 16 x (64*32) \\
 LeakyRELU & & 16 x (64*32) \\
 UpConv1DBlock & 25k, 4s, 64*16d & 64 x (64*16)\\
 UpConv1DBlock & 25k, 4s, 64*8d & 256 x (64*8)\\
 UpConv1DBlock & 25k, 4s, 64*4d & 1024 x (64*4)\\
 UpConv1DBlock & 25k, 4s, 64*2d & 4096 x (64*2)\\
 UpConv1DBlock & 25k, 2s, 1d & 8192 x 1\\
 
 \hline
\end{tabular}
\label{table:generator}
\end{table}

\begin{table} [ht]
\centering
\caption{Our discriminator network's structure. The DownConv1DBlock network consists of Conv1d (w/ stride) - Batch Norm - LeakyRELU - PhaseShuffle~\cite{wavegan} layers. The spectral normalization~\cite{miyato2018spectral} was applied to all layers. In the parameters column, k is the kernel length, s is the stride, d is the number of outputs.}
\begin{tabular}{ lll } 
 \hline
 Operation & Parameters & Outputs \\ 
 \hline
 Input(audio data, pitch label) & & 8192 x 1, 1\\
 DownConv1DBlock & 25k, 4s, 64d & 2048 x 64 \\
 DownConv1DBlock & 25k, 4s, 64*2d & 512 x (64*2)\\
 DownConv1DBlock & 25k, 4s, 64*4d & 128 x (64*4)\\
 DownConv1DBlock & 25k, 4s, 64*8d & 32 x (64*8)\\
 DownConv1DBlock & 25k, 2s, 64*16d & 16 x (64*16)\\
 Reduce Sum & axis=0 & 64*16 (save to 'x')\\
 Dense & 1d & 1 (save to 'output') \\
 \hline
 Projection~\cite{miyato2018cgans} & & \\
 \hline
 Input(pitch label) & & 1 \\
 Embedding & & 64*16 (save to 'y') \\
 Inner Product('x', 'y') & & 1 (save to 'projection') \\
 Add('output', 'projection') & & 1 \\
 \hline
\end{tabular}
\label{table:discriminator}
\end{table}

\subsubsection{Audio Embedding Network}
Our audio embedding network is consists of 5 DownConv1DBlock blocks as in the discriminator network, but with the PhaseSuffle layers removed and followed by a fully connected and a L2 normalization layers. The spectral normalization technique was not applied to this network. See Table~\ref{table:audio_embedding}.

\begin{table} [ht]
\centering
\caption{Our audio embedding network's structure. The DownConv1DBlock network consists of Conv1d (w/ stride) - Batch Norm - LeakyRELU layers. In the parameters column, k is the kernel length, s is the stride, d is the number of outputs.}
\begin{tabular}{ lll } 
 \hline
 Operation & Parameters & Outputs \\ 
 \hline
 Input(audio data) & & 8192 x 1\\
 DownConv1DBlock & 25k, 4s, 64d & 2048 x 64 \\
 DownConv1DBlock & 25k, 4s, 64*2d & 512 x (64*2)\\
 DownConv1DBlock & 25k, 4s, 64*4d & 128 x (64*4)\\
 DownConv1DBlock & 25k, 4s, 64*8d & 32 x (64*8)\\
 DownConv1DBlock & 25k, 2s, 64*16d & 16 x (64*16)\\
 Reduce Mean & axis=0 & 64*16 \\
 Dense & 1d & 1 \\
 L2-normalization & & 1\\
 \hline
\end{tabular}
\label{table:audio_embedding}
\end{table}

\subsubsection{Classification Network for IS and FID}
To compute the Inception Score (IS)~\cite{salimans2016improved} and Fr\^echet Inception Distance (FID)~\cite{heusel2017gans}, we trained classification networks that estimates the pitch or the instrument family labels on the NSynth training samples. The network exactly same as the audio embedding network except that the L2-nomrlaization layer is replaced with the softmax layer. 

\subsubsection{Auxiliary Discriminator Network}
The auxiliary discriminator consists of 5 fully connected layer - leaky RELU layer blocks and all the weight matrices are normalized by the spectral normalization technique. See Table~\ref{table:auxiliary_discriminator}. 

\begin{table} [ht]
\centering
\caption{Our auxiliary discriminator network's structure. In the parameters column, d is the number of outputs.}
\begin{tabular}{ lll } 
 \hline
 Operation & Parameters & Outputs \\ 
 \hline
 Input(audio feature) & & 505\\
 Dense & 128d & 128 \\
 LeakyRELU & & 128 \\
 Dense & 64d & 64 \\
 LeakyRELU & & 64 \\
 Dense & 32d & 32 \\
 LeakyRELU & & 32 \\
 Dense & 16d & 16 \\
 LeakyRELU & & 16 \\
 Dense & 1d & 1 \\
 \hline
\end{tabular}
\label{table:auxiliary_discriminator}
\end{table}

\subsubsection{Adversarial AutoEncoder}
We trained an adversarial autoencoder model proposed by ~\cite{makhzani2015adversarial} on the NSynth audio samples and modified it to enforce that the encoder outputs latent vectors that fit to the distribution given by VGGFace2 face feature vectors. The encoder and the decoder networks of the autoencoder have the same structure of the audio feature embedding network (Table~\ref{table:audio_embedding}) and the generator network (Table~\ref{table:generator}) described in the above, respectively, and the discriminator that is attached to the encoder's latent outputs has the same structure of the auxiliary discriminator (Table~\ref{table:auxiliary_discriminator}) described above. To get better quality of the audio sounds, we further enforce that the output audio of the decoder as close to real audio as possible by an additional discriminator whose structure is same as the one of the main discriminator (Table~\ref{table:discriminator}).

\subsection{Training Details}
The generator and the main and auxiliary discriminators are optimized with the standard adversarial losses and distance metric preservation loss as we discussed in Section 2. We used the Adam solver~\cite{kingma2014adam} for all models, with a learning rate of 0.0002 and momentum parameters $\beta_1$ = 0.5, $\beta_2$ = 0.999. We lowered the learning rate as the training progresses using an exponential decay function with 0.9 decay rate at every 100 epochs, $lr = lr * 0.9^{\frac{\text{epochs}}{\text{decay steps}}}$. We employed the early-stopping approach to stop our training by observing the optimization convergence. As in~\cite{miyato2018cgans}, we updated the two discriminators five times per each update of the generator. Our training mini-batch size is 128.

The audio feature network is optimized with the triple loss, and we set its margin to $1.0$ in our experiments. For more details we refer readers to \cite{schroff2015facenet}. The training mini-batch size is 128 and we stopped the training once we observed the triplet loss converged. 

The NSynth dataset contains about 300k four second monophonic 16kHz musical note samples, each with a unique pitch, timbre, and envelope. Those musical notes were synthesized or recorded from 10 different acoustic or electronic instruments; bass, brass, flute, guitar, keyboard, mallet, organ, reed, string, and vocal. In our experiments, we used only musical notes recorded from the acoustic instruments and whose pitch ranges from 24 to 84, and for each audio data, only the first slice of 8192 samples - i.e. about 0.5 sec length was used. As a result, total 60788 musical notes of the train split set defined in the NSynth dataset were used to train our models.